\documentstyle[11pt]{article}
\input epsf
\newlength{\awidth}
\newlength{\aheight}
\newlength{\uswidth}
\newlength{\usheight}
\newlength{\spacing}
\renewcommand{\arraystretch}{1.5}
\newlength{\topoff}
\newlength{\margoff}
\newlength{\margin}
\newlength{\hmargin}
\newcounter{fignr}
\newenvironment{fig}[1]{\refstepcounter{fignr}\label{#1}\begin{center}}{
    \end{center}}
\newcommand{\figcap}[2]{\parbox{#1}{{  Fig. \thefignr. #2}}}

\begin{document}
\setlength{\baselineskip}{\spacing}
\def\scri{
\unitlength=1.00mm
\thinlines
\begin{picture}(3.5,2.5)(3,3.8)
\put(4.9,5.12){\makebox(0,0)[cc]{$\cal J$}}
\end{picture}}

\begin{titlepage}
\noindent {May, 1998 \hfill USITP-98-06}\\[3 ex]
\begin{center}
{\Large A SPINNING ANTI-DE SITTER WORMHOLE}\\
\vspace{20mm}

{\large Stefan \AA minneborg}$^*$\footnote{Email address: 
stefan@vanosf.physto.se}

\vspace{5mm}

{\large Ingemar Bengtsson}$^{**}$\footnote{Email address: 
ingemar@vana.physto.se}
\vspace{5mm}

{\large S\"{o}ren Holst}$^{**}$\footnote{Email address: 
holst@vanosf.physto.se}

\vspace{14mm}

*{\sl \ Norra Reals Gymnasium, S-113 55 Stockholm, Sweden}

** {\sl \ Fysikum, Stockholm University, Box 6730, S-113 85 
Stockholm, Sweden}

\vspace{20mm}

{\bf Abstract}\\
\ \\
\end{center}
We construct a 2+1 dimensional spacetime of constant curvature 
whose spatial topology is that of a torus with one asymptotic 
region attached. It is also a black hole whose event horizon 
spins with respect to infinity. An observer entering the hole 
necessarily ends up at a "singularity"; there are no inner 
horizons. 
In the construction we take
the quotient of 2+1 dimensional anti-de 
Sitter space by a discrete group ${\Gamma}$. A key part of the 
analysis proceeds by studying the action of ${\Gamma}$ on the 
boundary of the spacetime.
\end{titlepage}
\newpage
\noindent {\bf 1. Introduction}\\[5 mm]
A topological geon is a gravitating object in which a 
non-trivial spatial topology is somehow "localized" in the interior. 
Geons have been much discussed over the years \cite{Wheeler}. 
By now a number of general results are available \cite{Gannon} 
but exact solutions remain hard to come by. In a previous 
publication (with Brill and Peld\'{a}n \cite{Fem}) 
we pointed out that if we restrict ourselves to 2+1 dimensional 
spacetimes then wormhole solutions of the vacuum equations are 
easy to construct, provided that a negative cosmological 
constant is admitted. Our definition of a wormhole is that it 
is a solution of Einstein's equations whose spatial topology 
is that of a non-simply connected surface with one asymptotic 
region attached. We 
constructed all such solutions under the restriction that they 
be time symmetric, and we analyzed how the general results 
apply to these spacetimes. The reason why these solutions 
are so easy to obtain is that in 2+1 dimensions all solutions 
of the vacuum equations have constant curvature; since we 
assumed a negative cosmological constant they are locally 
isometric to 2+1 dimensional anti-de Sitter space. It is 
believed \cite{Mess} (and we assume it to be true, also in 
the non-compact case) that all such solutions can be obtained 
as the quotient space
$ adS/{\Gamma} \ $,
where $adS$ denotes an open region of 2+1 dimensional 
anti-de Sitter space and ${\Gamma}$ is some discrete subgroup of 
isometries of this spacetime. Thus the question becomes how 
to identify those choices of discrete subgroups of isometries 
that lead to quotient spaces having the wormhole topology. 
An additional requirement is that all the "singularities" 
that occur to the future of some smooth spacelike slice must 
be hidden from view at infinity by an event horizon.

We are faced with a Lorentzian version of Clifford-Klein's 
problem \cite{Killing}: To determine all three dimensional 
spacetimes of constant negative curvature which are everywhere 
regular in the sense specified. 

We find this problem interesting in itself, and also because 
the resulting solutions can be used to test ideas about 
(say) topological censorship \cite{Friedman} or black hole 
entropy \cite{Carlip}. Unfortunately the present paper rests 
to a large extent on our previous work, and we found it 
difficult to make it self-contained while keeping its 
length within reasonable limits. For this reason 
section 2 provides a somewhat extended introduction; the 
organization of the rest of the paper is given away at the 
end of that section. Anyway, the key problem to be solved 
here is the construction of a wormhole whose event horizon 
spins with respect to infinity.

To get the priorities clear, it remains to say that our papers 
are basically an elaboration of results due to Ba\~{n}ados, 
Henneaux, Teitelboim and Zanelli \cite{BHTZ}.

\vspace{1cm}

\newpage
\noindent {\bf 2. The Introduction Continued}\\[5 mm]
The isometry group of 2+1 dimensional anti-de Sitter 
space is $SO(2,2)$, a group that obeys the local isomorphism 
\begin{equation} SO(2,2) \sim SL(2, {\bf R})\otimes SL(2, {\bf R}) 
\ . \end{equation}
The boundary of conformally compactified anti-de Sitter 
space is a 1+1 dimensional cylinder whose conformal structure is 
fixed, and the isometry group of the interior acts as the conformal 
group of the boundary. We refer to the boundary as \scri \ ("script 
I"). To construct our wormholes we first select a discrete subgroup 
${\Gamma}$ consisting of elements that can be reached by exponentiating 
some Killing vectors. The subgroup will be defined by listing a 
finite number of generators of ${\Gamma}$. As our covering 
space we choose an open region of anti-de Sitter space which 
is such that the flows of all the Killing vectors that generate 
elements of ${\Gamma}$ are spacelike within the region. This ensures 
that there are no closed timelike curves in the quotient space. 
The boundary of covering space 
is considered to give rise to "singularities" in the quotient 
space (and we use quotation marks since it may to some extent 
be possible to analytically extend through the "singularities" 
at the cost of admitting closed timelike curves---this notion of 
"singularity" has been fully discussed in earlier papers 
\cite{BHTZ}, to which the sceptical reader is referred). 
The causal structure of the quotient space is most conveniently 
analyzed directly in covering space \cite{Tre}. Now a fully 
explicit description of the covering space needs a fully explicit 
description of all the infinite elements of ${\Gamma}$, and 
in general this is very hard to obtain. However, a description 
that is sufficiently explicit for our purposes needs only a 
sufficient amount of control over the location of those 
fixed points of the group that do occur on \scri . 

In our previous publication \cite{Fem} we assumed time symmetry, or in other 
words that ${\Gamma}$ is a subgroup of a diagonal $SL(2, {\bf R})$ 
subgroup that transforms a particular spacelike slice onto itself. 
This slice is a Poincar\'{e} disk, and the action of the diagonal 
subgroup is that of M\"obius transformations preserving the disk. 
It is then possible to solve the problem in two steps \cite{Brill}: 
The group ${\Gamma}$ transforms the Poincar\'{e} disk 
onto itself and therefore one can begin by choosing ${\Gamma}$ 
so that it gives the appropriate smooth spatial topology to this 
surface. Once this has been ensured the action of ${\Gamma}$ is extended 
to a suitable open region of the full anti-de Sitter space and 
the resulting spacetime analyzed in detail. In the first step we 
have to ensure that none of the infinite set of elements of ${\Gamma}$ 
have fixed points within the disk. In group theoretical terms, this 
requirement translates into the statement that all the elements of 
${\Gamma}$ have to belong to the hyperbolic conjugacy class. 
Hyperbolic elements always have a pair of fixed points lying on 
the boundary of the disk, but this is allowed. In practice ${\Gamma}$ 
will be defined by choosing a fundamental region that tessellates 
the disk. By specifying which pairs of edges of the fundamental 
region that are to be identified one specifies a finite set 
of generators of the group ${\Gamma}$, and the condition on 
the rest of the elements is automatically met provided that 
the manifold that one obtains when gluing the edges of the 
fundamental region together is smooth. Fixed points do occur 
on the boundary of the disk, and it is not too 
difficult to pin them down with the accuracy that one needs. 
(Details will be given later.) Thus at the moment of time 
symmetry the covering space of the interior is the entire disk, 
and the covering space of the boundary is all of the boundary 
except for the fixed points. In the next step of the construction 
we must first define the boundaries of the covering space of 
our spacetime. According to our definition \cite{Tre} these 
boundaries are the "singularity surfaces" where a Killing 
vector generating some element of ${\Gamma}$ becomes lightlike. 
When ${\Gamma}$ belongs to the diagonal subgroup of $SO(2,2)$ 
it happens that the "singularity surfaces" are null planes, that 
is to say light cones with their vertices on \scri . These 
vertices are precisely the fixed points that we encountered in 
the first step. In this way we can understand our covering 
space. Additional fixed points occur where the null planes 
intersect to the future (and to the past) of our initial data 
slice. This is where spacetime, and \scri \ , ends. Only a part of 
spacetime can be seen from infinity, and in this sense our 
wormhole turns out to be a {\it bona fide} black hole. The event 
horizon is the backwards light cone of the last 
point on \scri . (Again, details will be given later.) 

Now time symmetry is a strong restriction that we want to lift. 
The restriction was imposed only because it facilitates finding 
the location of some of the fixed points on \scri . However, the 
same information can be found in a different way. The point---easy to 
see for those readers who are familiar with conformal field 
theory---is that if we use light cone coordinates $u$ and $v$ 
on \scri , then the action of the two $SL(2, {\bf R})$ factors of 
$SO(2,2)$ decouples. One factor will give projective 
transformations on the lines defined by constant $v$, and 
the other on the lines defined by constant $u$. This 
observation is enough to enable us to formulate conditions 
on the generators of ${\Gamma}$ so that the quotient of 
any region of \scri \ where the Killing vectors are spacelike 
becomes a smooth manifold terminated by a "singularity" caused 
by fixed points. In other words, in the time symmetric case 
we were able to start off our discussion by first making a 
sufficiently accurate picture of the covering space of the initial 
data slice, but we now see that we can instead start from a 
sufficiently accurate picture of the covering space of \scri . 
In the next step the analysis proceeds as in the time symmetric 
case, although there are some differences. For one thing the 
event horizon now turns out to "spin" relative to \scri \ . 
There is also a definite difficulty caused by the fact that 
the "singularity surfaces" are no longer null planes, even 
though their intersections with \scri \ are null \cite{Tre}. 
In the case 
of the Kerr-like BTZ black hole \cite{BHTZ} what happens is 
that the covering space increases as the spin of the event 
horizon increases, with the result that a "mouth" opens up in the 
interior through which it is possible to travel to another 
universe. Our analysis shows that this is not the case for 
the spinning wormhole. The interior of a wormhole always ends 
in a "singularity", and there are no inner horizons.

There is a rather large amount of formul\ae \  in the body of the 
paper. This is because we decided to include enough detail so 
that we could sketch an explicit calculation of the angular 
velocity of the horizon in terms of the parameters characterizing 
the generators of ${\Gamma}$. It seemed worthwhile to do this 
even though it would be possible to understand the construction 
with less attention to detail.

After this outline of the argument, we are ready to state 
the contents of our paper: Section 3 provides the necessary 
background information about anti-de Sitter space and its 
isometry group. Section 4 gives an account of the BTZ black holes 
\cite{BHTZ}. Although this is standard material our presentation 
is new, and it is intended to show how information may be 
extracted from \scri . In section 5 we revisit the spinless 
wormholes \cite{Fem}, but again from a new point of view that 
stresses the properties of the covering space. In section 6 
we finally prove that a spinning wormhole exists, 
we compute the angular velocity of its horizon, and we show that 
its domain of exterior communication is isometric to that of 
a BTZ black hole. We also count the number of parameters in 
our solution. In section 7 we discuss the interior of the 
spinning wormhole and prove that it ends in a "singularity"; 
there is only one asymptotic region and---according to an 
admittedly somewhat heuristic argument---there are no inner 
horizons. Finally, section 8 states our conclusions and 
provides a list of open questions.

Although we have made an effort to write a readable paper it 
will help if the reader has at least a nodding acquaintance 
with our previous publications \cite{Fem} \cite{Tre}, as well 
as with standard definitions in black hole physics \cite{Wald}. \\[1 cm]
{\bf 3. 2+1 anti-de Sitter space and its isometry group.}\\[5 mm]
Three dimensional anti-de Sitter space can be defined 
as the quadric surface
\begin{equation} X^2 + Y^2 - U^2 - V^2 = - 1 \end{equation}
 embedded in a flat space with the metric
\begin{equation} ds^2 = dX^2 + dY^2 - dU^2 - dV^2 \ . \end{equation}
 A description of anti-de Sitter space in terms 
of embedding coordinates is not quite enough for our purposes 
since we will want to define the conformal boundary 
of spacetime. Therefore we introduce the 
intrinsic coordinates $t$, ${\rho}$, ${\phi}$ through
\begin{equation} 
X = \frac{2{\rho}}{1 - {\rho}^2}\cos{\phi} \hspace{3 ex} Y = \frac{2{\rho}}{1 - {\rho}^2}\sin{\phi}
\end{equation}
\begin{equation} 
U  = \frac{1 + {\rho}^2}{1 - {\rho}^2}\cos{t} \hspace{3 ex} V = \frac{1 + {\rho}^2}{1 - {\rho}^2}\sin{t}  . 
\end{equation}
 Then (the universal covering space of) anti-de Sitter 
space becomes simply the interior of the (infinite) cylinder 
${\rho} < 1$, and its conformal compactification is obtained 
by adjoining the surface of the cylinder ${\rho} = 1$. This 
boundary is \scri , the set of endpoints of all lightlike geodesics. 
Note that past \scri , future \scri , and spacelike infinity all 
coincide for anti-de Sitter space. The asymptotic structure 
is therefore quite different from that of Minkowski space 
(and it is this very feature that enables one to find black 
holes with constant curvature). The metric on anti-de 
Sitter space becomes in our coordinates
\begin{equation} ds^2 = - \left( \frac{1 + {\rho}^2}
{1 - {\rho}^2}\right)^2dt^2 + \frac{4}
{(1 - {\rho}^2)^2}(d{\rho}^2 + {\rho}^2d{\phi}^2) 
\ . \end{equation}
 At constant $t$ this is the metric for the 
hyperbolic plane, also known as the Poincar\'{e} disk. 

A metric on \scri \ can be obtained by multiplying 
the metric of anti-de Sitter space with a suitable factor 
that vanishes on \scri. In this way we obtain an "unphysical" 
metric that extends smoothly to the boundary, and it will induce 
a metric there. Our choice for the unphysical metric is 
\begin{equation} d\hat{s}^2 = \frac{1}{U^2 + V^2}ds^2 
= - dt^2 + \frac{4}{(1 + {\rho}^2)^2}
(d{\rho}^2 + {\rho}^2d{\phi}^2)\ . \end{equation}
 This metric induces the flat metric 
on the cylindrical surface ${\rho} = 1$. It will prove convenient 
to introduce light cone coordinates on the boundary; 
\begin{equation} 
\label{nullcoord}
u = t - {\phi} \hspace{15mm} v = t + {\phi} \ . \end{equation}
 In terms of these coordinates the metric on \scri \ 
takes the form 
\begin{equation} d\hat{s}^2 = - dudv \ . \end{equation} 
 For a discussion of the conformal compactification of 
asymptotically anti-de Sitter spaces in general, see ref. \cite{Ashtekar}.

The group of isometries of three dimensional anti-de Sitter space 
is the six dimensional group $O(2,2)$. Its connected 
component $SO_o(2,2)$ is a direct product 
\begin{equation} SO_o(2,2) = \frac{SL(2,{\bf R})\otimes 
SL(2,{\bf R})}{Z^2} \ . \end{equation}\\[.5 ex]
To see this, write the defining equation of the 
quadric as a condition on the determinant of a matrix: 
\begin{equation} |{\bf X}| = \left| \begin{array}{cc} V+X & Y+U \\ 
Y-U & V-X \end{array} \right|  = 1 \ . \end{equation}
 This condition is clearly preserved 
by any transformation of the form
\begin{equation} {\bf X} \rightarrow {\bf X}' = g{\bf X}\tilde{g}^{-1} 
\ , \hspace{1cm} g \in SL(2, {\bf R}) \ , \hspace{6mm} \tilde{g} 
\in SL(2, {\bf R}) \ . \end{equation}
 In this way any element $G$ of $SO_0(2,2)$ may be 
identified with an equivalence class of two elements in the direct 
product of two $SL(2, {\bf R})$s; 
\begin{equation} G = (g, \tilde{g}) \sim (-g, -\tilde{g}) \ . \end{equation}
 Group multiplication is defined in the obvious way. An 
example of an element of $SO(2,2)$ that does not lie 
in the connected component is the reflection 
\begin{equation} {\pi} \ : \hspace{1cm} 
(X,Y,U,V) \rightarrow (X,-Y,U,-V) \ . \end{equation}
 An example of an element of $O(2,2)$ that does not lie 
in $SO(2,2)$ is the reflection 
\begin{equation} {\Pi} \ : \hspace{1cm} 
(X,Y,U,V) \rightarrow (X,Y,U,-V) \ . \end{equation}
 Once the structure of $SL(2,{\bf R})$ 
has been understood an understanding of $O(2,2)$ is not far behind. 

The group $SL(2, {\bf R})$ is defined as the group formed by 
all two by two matrices of unit determinant. Let us coordinatize an 
arbitrary element $g \in SL(2,{\bf R})$ as 
\begin{equation} g = \left( \begin{array}{cc} D + A & B + C \\ 
B - C & D - A \end{array} \right) \ , \hspace{5mm} |g| = 1 \ . \end{equation}
 We see that the group manifold of $SL(2,{\bf R})$ is 
precisely 2+1 dimensional anti-de Sitter space (a happy accident!). 
Hence we can draw a Penrose diagram of our group, see figure~\ref{fig1}.
It is naturally divided into regions by the lightcones having vertices at the 
group elements $\pm e$. If we identify group elements that differ 
by a sign we obtain $SO_0(2,1)$, the group of 
M\"{o}bius transformations that preserve the unit circle and 
map its interior to itself. Alternatively we can describe 
$SO_0(2,1)$ as the group of projective transformations of 
the circle. Both viewpoints will figure prominently below.

We will be interested in understanding the conjugacy classes 
of the group, and we will want to know which 
group elements can be reached by exponentiating its Lie algebra 
elements. Let us begin with the conjugacy classes of 
$SL(2, {\bf R})$, which are 
equivalence classes of elements that can be connected to each 
other by means of conjugation with some group element $g_0$; 
\begin{equation} g \sim g' = g_0gg_0^{-1} \ . \end{equation}
 The trace of $g$ is invariant under conjugation. The 
conjugacy classes are surfaces that foliate the group, and in 
our coordinates it is easy to see how: 
\begin{equation} \mbox{Tr}\ g = 2D \hspace{5mm} \Rightarrow 
\hspace{5mm} A^2 + B^2 - C^2 = - 1 + D^2 = - 1 + 
\left( \frac{\mbox{Tr}\ g}{2}\right)^2 \ . \end{equation}
In this way we find that the various regions of our 
Penrose diagram correspond to conjugacy classes according to 
the following scheme:
\setlength{\arraycolsep}{3 ex}
\renewcommand{\arraystretch}{2}
$$
\begin{array}{lll}
\mbox{Tr}\ g > 2   & \mbox{Region I} & \mbox{(hyperbolic)} \\
\mbox{Tr}\ g < - 2 & \mbox{Region II} & \mbox{(hyperbolic)} \\
|\mbox{Tr}\ g| < 2 & \mbox{Regions III, IV} & \mbox{(elliptic)} \\
|\mbox{Tr}\ g| = 2 & \mbox{The light cones} & \mbox{(parabolic)} . 
\end{array}
$$
The notation "hyperbolic, elliptic and parabolic" 
refers to the classification of M\"{o}bius transformations.\\
% Fig 1:
\parbox{\textwidth}{\begin{fig}{fig1}
  \mbox{\setlength{\epsfysize}{.40\textheight} \epsfbox{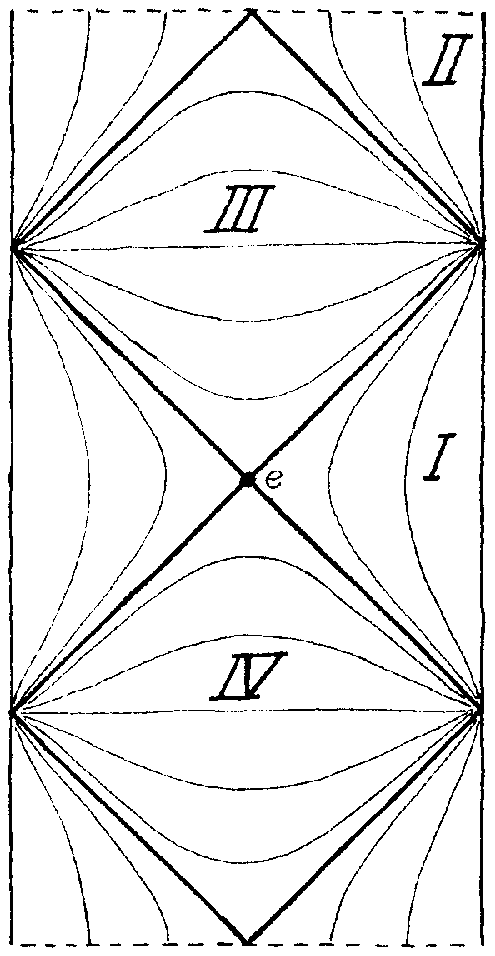} }\\[2 ex]
  \figcap{.9\textwidth}{ The Penrose diagram of $SL(2, {\bf R})$. 
The lightcones shown naturally divide it into regions, and 
these have been foliated by conjugacy classes. 
Note that region II cannot be reached by exponentiation from the unit 
element $e$.}
\end{fig}
}\\[2 ex]
\indent If we exponentiate a Lie algebra element we obtain a 
geodesic in the group manifold that starts out from the unit 
element $e$. From this we may deduce that region II cannot 
be reached by exponentiation in this way. The exponentiation 
will be done explicitly in section 6, where we will regard 
the group as generating projective transformations of the 
circle. 

We now turn to $O(2,2)$, whose connected component is a product 
of two three dimensional anti-de Sitter spaces quotiented 
by $Z^2$. Our interest lies in discrete subgroups of $O(2,2)$ 
whose generators can be reached by exponentiation from the 
identity (this is a restriction that we impose somewhat arbitrarily), 
and our task is therefore to divide the connected 
component into conjugacy classes with respect to the full 
group $O(2,2)$. Given our knowledge of $SL(2, {\bf R})$ 
we already understand conjugation with elements in the 
connected component (because it can be carried out 
separately on the two factors of the group). Conjugation 
with the reflection ${\Pi}$ that was defined above has the 
effect of exchanging the two factors; 
\begin{equation} {\Pi}(g_1, g_2){\Pi}^{-1} = (g_2, g_1) 
\ . \end{equation}
 With this bit of additional information the problem 
is solved, and we find the following conjugacy classes (the 
result and the notation is due to Ba\~nados et al. \cite{BHTZ}):

\

$I_a$: \hspace{12mm} elliptic $\otimes$ hyperbolic 

$I_b$: \hspace{12mm} hyperbolic $\otimes$ hyperbolic 

$I_c$: \hspace{12mm} elliptic $\otimes$ elliptic

$II_a$: \hspace{1cm} parabolic $\otimes$ hyperbolic 

$II_b$: \hspace{1cm} parabolic $\otimes$ elliptic 

$III_a$: \hspace{8mm} parabolic $\otimes$ parabolic, future $\otimes$ past

$III_b$: \hspace{8mm} parabolic $\otimes$ parabolic, future $\otimes$ future.
\\[2 ex]
As it turns out the only class of interest to us is 
type $I_b$. We will be interested in formulating conditions that 
guarantee that all the elements of a discrete subgroup ${\Gamma}$ 
lie in this class. Why this is the problem will begin to be 
apparent in the next section.

We end this section with a list of the Killing vectors of anti-de 
Sitter space. In embedding coordinates they are 
\begin{equation} J_{XY} = X\partial_Y - Y\partial_X \hspace{1cm} 
J_{XU} = X\partial_U + U\partial_X \end{equation} 
 and so on. Since the Lie algebra of $SO(2,2)$ splits 
in a direct sum it is convenient to group them into two mutually 
commuting sets. We will also need to know how the Killing vectors 
act on \scri ; here is a list that accomplishes both purposes:
\begin{equation} J_1 \equiv - \frac{1}{2}(J_{XU} + J_{YV}) = 
\sin{u}\partial_u \hspace{5mm} \tilde{J}_1 \equiv - \frac{1}{2}(J_{XU} 
- J_{YV}) = \sin{v}\partial_v \end{equation}
\begin{equation} J_2 \equiv - \frac{1}{2}(J_{XV} - J_{YU}) = - 
\cos{u}\partial_u \hspace{5mm} \tilde{J}_2 \equiv - \frac{1}{2}(J_{XV} 
+ J_{YU}) = - \cos{v}\partial_v \end{equation}
\begin{equation} J_3 \equiv - \frac{1}{2}(J_{XY} - J_{UV}) = 
\partial_u \hspace{12mm} \tilde{J}_3 \equiv \frac{1}{2}(J_{XY} 
+ J_{UV}) = \partial_v \ . \hspace{5mm} \end{equation} 
 Admittedly there is some abuse of notation here---the 
expressions in terms of embedding coordinates are only valid in 
the interior, and the expressions in light cone coordinates only 
on the boundary---but we hope that the meaning is clear. As promised, 
the use of light cone coordinates on \scri \ makes
the split into two mutually commuting sets manifest. Note 
that on \scri \ these vector fields are in general only conformal 
Killing vectors with respect to the metric that we introduced on 
\scri . But then it is only the conformal structure on \scri \ that 
has an invariant significance. As a side remark we observe that 
the conformal group in two dimensions is infinite dimensional, 
and that in fact the entire conformal group acts as an asymptotic 
symmetry group here \cite{Brown}. But this fact will play no role 
in the present paper. 

The classification of the group elements into conjugacy 
classes applies to the Killing vectors as well. The general form of 
a Killing vector that belongs to 
the hyperbolic conjugacy class of $SL(2, {\bf R})$ is 
\begin{equation} {\xi} = x_1J_1 + x_2J_2 + x_3J_3 \ , \hspace{1cm} 
x_1^2 + x_2^2 - x_3^2 > 0 \ . \end{equation}
Up to normalization we can write this in the alternative form 
\begin{equation} {\xi} = \sin{\alpha}J_1 - \cos{\alpha}J_2 
- \cos{\beta}J_3 \ , \hspace{1cm} {\beta} \neq 0 \ . \end{equation}
The fact that an arbitrary arbitrary hyperbolic 
Killing vector can be written like this happens to be useful later. 
And with this observation our group 
theoretical preliminaries are finally at an end.\\[1 cm]
{\bf 4. The BTZ Black Hole Viewed From \scri .}\\[5 mm]
As shown by Ba\~nados, Henneaux, Teitelboim and 
Zanelli \cite{BHTZ} a black hole solution of Einstein's equations 
can be obtained by choosing a Killing vector from the conjugacy 
class $I_b$, letting it generate a finite group element ${\gamma}$, 
and taking the spacetime to be $adS/{\Gamma}$, where $adS$ is a 
region where the flow of the Killing vector is spacelike and 
${\Gamma}$ is the cyclic group generated by ${\gamma}$. The solution 
has many properties in common with the Kerr family of black holes 
in four spacetime dimensions. The reason why the conjugacy class 
$I_b$ is singled out here has to do with the location of the fixed 
points and the "singularity surfaces" (where the generating Killing 
vector becomes lightlike); a geometrical analysis of this question, 
valid in 2+1 and 3+1 dimensions, can be found in the literature \cite{Tre}. 
In 2+1 dimensions there is an extremal black hole that results from 
choosing a Killing vector of type $III_a$, but it will play no role 
in the present paper. 

We will present the BTZ black hole from a new point of view that 
will prove useful in the following sections. The construction starts 
by selecting a group element generated by a Killing vector of type 
$I_b$, say 
\begin{equation} {\gamma}_1 = e^{{\xi}_1} \ , \hspace{5mm} 
{\xi}_1 = aJ_{XU} + bJ_{YV} = - (a+b)J_1 - 
(a-b)\tilde{J}_1 \ . \end{equation}
 Without loss of essential generality we may choose 
$a > b \geq 0$. Let us consider the flow of this Killing vector 
on \scri . The flow will be spacelike in the square defined by 
(say) 
\begin{equation} - {\pi} < u < 0 \hspace{12mm} 0 < v < {\pi} 
\ . \end{equation} 
 This region will be the covering space of an asymptotic 
region of the BTZ black hole. To show that the BTZ black hole 
is asymptotically anti-de Sitter we introduce a new metric 
on \scri :
\begin{equation} d\hat{\hat{s}}^2 = - \frac{dudv}{(a^2 - b^2)\sin{u}\sin{v}} 
\ . \end{equation}
 This metric is related by a conformal rescaling to 
the one introduced above and the transformation is well defined 
in the entire covering space, so this is allowed. Our Killing vector 
${\xi}_1$ is indeed a true Killing vector of this new metric. When the 
identification is carried through, the Killing vector ${\xi}_1$ will 
have closed flow lines and can therefore be regarded as the 
generator of rotations in the quotient space, which has the topology 
of a cylinder. The quotient space also admits a global Killing vector 
orthogonal to the generator of rotations. It has a timelike flow 
and can therefore be regarded as the generator of time translations. 
The boundary of the quotient space has now been shown to be 
a timelike cylinder with the appropriate conformal 
structure---hence the BTZ black hole is asymptotically anti-de 
Sitter. To be explicit about it, the 
generator of asymptotic rotations is
\begin{equation} {\xi}_{rot} \equiv {\xi}_1 = - (a+b)\sin{u}\partial_u - 
(a-b)\sin{v}\partial_v \ . \end{equation}
 Up to normalization the generator of time translations is 
determined by the condition
\begin{equation} {\xi}_{rot}\cdot {\xi}_{time} = 0 \ . \end{equation} 
 A solution is
\begin{equation} {\xi}_{time} = - (a+b)\sin{u}\partial_u + 
(a-b)\sin{v}\partial_v \ . \end{equation}
When evaluated with respect to the new metric on 
\scri \ the norms of these generators are
\begin{equation} ||{\xi}_{rot}||_{\infty}^2 = - ||{\xi}_{time}||_{\infty}^2 
= 1 \ . \end{equation} 
It is useful to have a rough idea about the flow 
of these Killing vectors for various values of the parameter 
$b/a$, see figure~\ref{fig2}.

What about the spin of the BTZ black hole? To compute it we 
need to know that the covering space of the event horizon is 
the backwards light cone of the last point on \scri, whose 
light cone coordinates are 
\begin{equation} (u,v) = (u_P, v_P) = (0, {\pi})\ . \end{equation}
%
% Fig 2:
\parbox{\textwidth}{\begin{fig}{fig2}
\parbox[b]{.05\textwidth}{
a)\\[.235\textheight]
b)\\[.24\textheight]
c)\\
\mbox{}
}
\mbox{\setlength{\epsfysize}{.75\textheight} \epsfbox{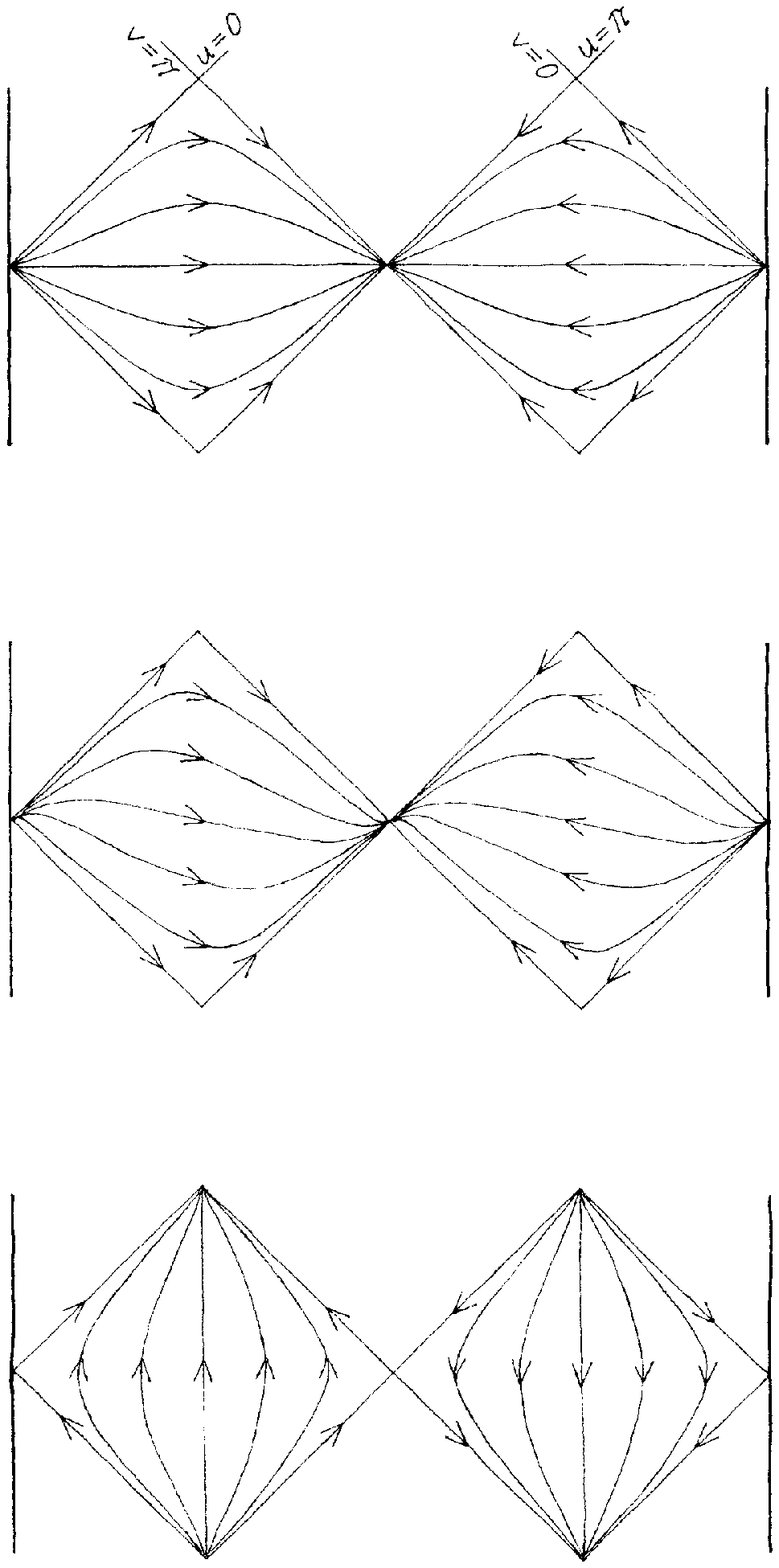} }\\[4 ex]
  \figcap{.98\textwidth}{  These pictures all show various Killing 
flows on the surface of the anti-de Sitter cylinder---that is, on
\scri---which has been cut along $\phi = 0$ and flattened out.
The null coordinates $u$ and $v$ defined in eqn. (\ref{nullcoord}) are
indicated in a). The flows are drawn only in the regions belonging to
the covering space of the BTZ black hole, that is, where the Killing
flow $\xi_1$ generating the identification is spacelike. \\
a) $\xi_1$ (or $\xi_{rot}$) for the spinless BTZ hole ($b=0$).\\
b) $\xi_1$ (or $\xi_{rot}$) for the spinning BTZ hole: ${\Omega} \equiv b/a 
= 1/4$.\\
c) The flow $\xi_{hor}$ generating the horizon. This flow is
independent of the spin, and it is easy to see from these pictures
when the black hole spins (cf. eqn. (\ref{xihor})).
}
\end{fig}
}
\newpage
\noindent In embedding coordinates a light cone with its vertex 
on \scri \ , say at $(t, {\phi}) = (t_P, {\phi}_P)$, is given by the equation
\begin{equation} \cos{{\phi}_P}X + \sin{{\phi}_P}Y - \cos{t_P}U 
- \sin{t_P}V = 0 \ . \end{equation}
 The formula holds for arbitrary points on \scri . Adapting it 
to our case we find that the union of the backwards lightcone 
of the last point on \scri \ with the forwards lightcone of the first 
point on \scri \ is given by 
\begin{equation} (Y - V)(Y + V) = Y^2 - V^2 = 0 \ . \end{equation} 
 This is a bifurcate surface and it is lightlike since it 
contains its own normal; the normal coincides with the Killing vector 
\begin{equation} {\xi}_{hor} = J_{YV} \ . \end{equation}
 Therefore the event horizon lies in the bifurcate Killing 
horizon ruled by the integral curves of this Killing vector, to which 
we refer as the horizon generator. Now we can bring the horizon 
generator back to \scri , where it can be written as a linear combination
\begin{equation} 
\label{xihor}
{\xi}_{hor} = - \sin{u}\partial_u + \sin{v}\partial_v 
= \frac{a}{a^2 - b^2}( {\xi}_{time} - {\Omega}{\xi}_{rot}) \ , \end{equation}
 where 
\begin{equation} {\Omega} = \frac{b}{a} \ . \end{equation}
 The overall normalization is not relevant. On the other 
hand the parameter ${\Omega}$ is a measure of how much rotation the 
horizon generator contains. By a definition that applies to 
any stationary black hole ${\Omega}$ is the angular velocity of 
the event horizon. 

Again by definition, the domain of exterior communication of a black hole 
contains all points that can be reached from \scri \ by both 
future and past directed causal curves. In our case it is bounded 
on one side by \scri \ and on the other by the bifurcate Killing 
horizon that we found. Until section 7 our interest will be 
entirely confined to the domain of exterior communication for all 
the black holes that we will construct.

By the way, there is one flow line of ${\xi}_1$ that is a spacelike 
geodesic, namely the line $Y = V = 0$. When the identification is 
carried through this becomes a closed geodesic of length $a$. But 
the intersection of the event horizon (a null plane in covering 
space) with the Poincar\'{e} disk defined by $V =0$ is also a 
geodesic, since both surfaces are totally geodesic. It is the same 
geodesic defined in two different ways; hence the length of the event 
horizon---which is the entropy up to a numerical factor---is 
just $a$. We also observe that the mass and the spin 
of the black hole can be expressed as functions of $a$ and $b$, but 
the definition really requires the kind of careful attention 
given to it in ref. \cite{BHTZ}. For our purposes the angular 
velocity is enough.

We will now repeat the BTZ construction for an arbitrary Killing 
vector belonging to type $I_b$. This may seem to be a pointless 
exercise since any such Killing vector can be brought to the form 
already considered by means of conjugation. We do it because in 
our discussion of the wormhole we will want to consider generating 
Killing vectors "in arbitrary position", and then the formul\ae \ 
that we obtain now will prove useful. So, by the remark at the end 
of the previous section, on \scri \ an arbitrary type $I_b$ Killing 
vector can always be written (up to normalization) in the form
\begin{eqnarray} {\xi} = - \frac{1}{2}(\sin{\alpha}\sin{u} 
+ \cos{\alpha}\cos{u} - \cos{\beta})\partial_u \nonumber \\
\ \label{xi} \\
+ \frac{k}{2}(\sin{\tilde{\alpha}}\sin{v} 
+ \cos{\tilde{\alpha}}\cos{v} - \cos{\tilde{\beta}})\partial_v \ . 
\nonumber \end{eqnarray}
 There are five real parameters. We can set 
\begin{equation} {\alpha} \equiv \frac{u_P + u_{P'}}{2} \hspace{5mm} 
{\beta} \equiv \frac{u_P - u_{P'}}{2} \hspace{5mm} \tilde{\alpha} \equiv 
\frac{v_P + v_{P'}}{2} \hspace{5mm} \tilde{\beta} \equiv 
\frac{v_P - v_{P'}}{2} \ , \end{equation} 
 and without loss of essential 
generality we can take $u_P > u_{P'}$, $v_P > v_{P'}$ and $k > 0$. The 
apparently somewhat eccentric choice of these parameters is 
explained when we observe that trigonometric identities can be 
employed to rewrite the Killing vector in the convenient form 
\begin{equation} 
\label{genxi}
{\xi} = \sin{\left( \frac{u - u_P}{2}\right) }
\sin{\left( \frac{u - u_{P'}}{2}\right) }
\partial_u - k\sin{\left( \frac{v - v_P}{2}\right) }
\sin{\left( \frac{v - v_{P'}}{2}\right) }\partial_v \ . 
\end{equation}
 The advantage is that now the general nature of the 
flow of the Killing vector on \scri \ is evident (figure~\ref{fig3}). In particular 
the flow will be lightlike 
along the lightlike lines $u = u_P$ or $u_{P'}$, $v = v_P$ or $v_{P'}$. 
The parameter $k$ is an additional parameter not determined by 
the location of the fixed points of the flow, and will---as we 
will see---enter into the formula for the spin of the BTZ black hole. 

We can now perform a conformal rescaling on a region of 
\scri \ where the flow is spacelike such that the norm of our 
chosen Killing vector becomes unity. To be precise about it, 
we choose 
\begin{equation} d\hat{\hat{s}}^2 = {\omega}^{-1}d\hat{s}^2 \ , \end{equation}
 where the conformal factor is
\begin{equation} {\omega} = - k \sin{\left( \frac{u - u_P}{2}\right) }
\sin{\left( \frac{u - u_{P'}}{2}\right) }\sin{\left( \frac{v - v_P}{2}\right) }
\sin{\left( \frac{v - v_{P'}}{2}\right) } \ . \end{equation} 
 When the identification is carried through the Killing 
vector ${\xi}$, whose norm on \scri \ is now unity, will serve as the 
generator of rotations. There is an orthogonal timelike Killing vector 
with unit norm that will serve as the generator of time translations. 
The event horizon is the backwards light cone of the last point 
on \scri \ , as usual. Fortunately it is not really necessary to 
go into spacetime to fetch its generator. It can be found by means 
of a calculation carried through entirely on \scri , as the Killing 
vector field that is normal to the surface 
\begin{equation} {\omega} = 0 \ . \end{equation}
%
% Fig 3:
\parbox{\textwidth}{\begin{fig}{fig3}
  \mbox{\setlength{\epsfxsize}{.70\textwidth} \epsfbox{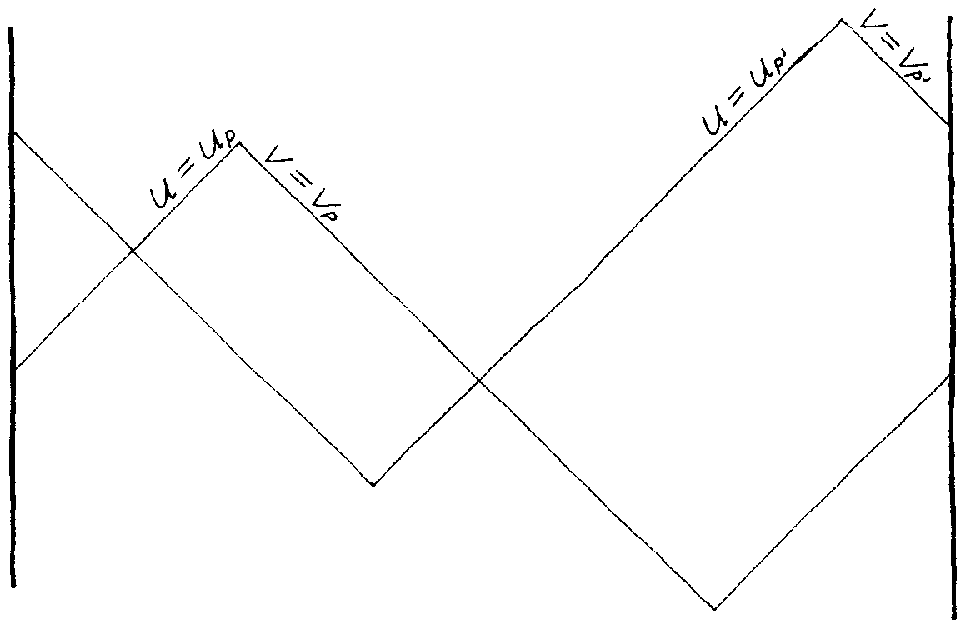} }\\[2 ex]
  \figcap{0.9\textwidth}{  This figure again is drawn on the surface
of the anti-de Sitter cylinder. The null lines are the Killing
horizons for a type $I_b$ Killing field ``in general position'' (eqn.
(\ref{genxi})). The flow is spacelike within the two rectangular
regions and has fixed points where these lines intersect. }
\end{fig}
}\\[2 ex]
\indent Finally we express the horizon generator as a linear 
combination of the generators of time translation and rotation, 
and read off the angular velocity of the event horizon relative 
to \scri. 
To cut a fairly long story short, the result is 
\begin{equation} {\Omega} = \frac{\sin{\left( \frac{u_P - u_{P'}}{2}\right)} - 
k\sin{\left( \frac{v_P - v_{P'}}{2}\right)}}
{\sin{\left( \frac{u_P - u_{P'}}{2}\right)} + 
k\sin{\left( \frac{v_P - v_{P'}}{2}\right)}} \ , \label{Omega} \end{equation}
 where $(u_P, v_P)$ are the light cone coordinates of 
the last point on \scri \ and $(u_{P'}, v_{P'})$ those of the 
first point. As in the special case considered earlier, the 
horizon generator will be the normal of a bifurcate Killing 
horizon that forms the inwards boundary of the domain of 
exterior communication of the black hole. This observation 
as well as the formula for ${\Omega}$ will prove useful below.\\[1 cm]
\newpage
{\bf 5. The Exterior of a Spinless Wormhole.}\\[5 mm]
The next and final stage in our preparations for the 
spinning wormhole is to revisit the spinless wormhole.  
Again the task is to select some discrete subgroup ${\Gamma}$ 
of isometries of anti-de Sitter space and then to take the quotient 
of a suitable open region of this space with ${\Gamma}$. The 
quotient space will necessarily have constant curvature; the 
difficult part is to ensure that its spatial topology is (say) 
a torus with a disk cut out, that \scri \ exists in the quotient 
space and that the "singularities" in the interior are hidden by 
an event horizon. The wormhole will be a spinless black hole 
provided that all the elements of ${\Gamma}$ belong to the diagonal 
subgroup $SO(2,1)$ of $SO(2,2)$ consisting of group elements 
that transform the Poincar\'{e} disk $V = 0$ onto itself, and we 
have to arrange matters so that only hyperbolic group elements 
occur. In our previous publication \cite{Fem} we solved the 
entire problem through the specification of a fundamental region 
that defines ${\Gamma}$. It turns out that this approach is 
impracticable in the spinning case, and therefore we will now 
present the spinless wormhole in covering space. This is 
enough to discuss the causal structure of the solution. 

We will concentrate on a simple and quite symmetric example, leaving 
the general case to take care of itself. Thus we choose the type 
$I_b$ group elements 
\begin{equation} {\gamma}_1 = e^{{\xi}_1} \ , \hspace{10mm} 
{\xi}_1 = aJ_{XU} \end{equation}
\begin{equation} {\gamma}_2 = e^{{\xi}_2} \ , \hspace{10mm} 
{\xi}_2 = aJ_{YU} \ . \end{equation}
 We assume that $a > 0$. The second Killing vector is 
obtained through a 90 degree rotation of the first. Now we take 
our discrete group ${\Gamma}$ to be the free group that is formed 
by all possible "words" constructed from these generators.
This group has an infinite set of elements, and it would 
be a very hard task to obtain an explicit characterization of all 
of them. In order to understand the action of ${\Gamma}$ on anti-de 
Sitter space it will therefore be necessary to proceed somewhat 
indirectly. 

The main point at issue is to find conditions on the parameter 
$a$ such that all the elements of ${\Gamma}$ can be generated by type 
$I_b$ Killing vectors. The generators of ${\Gamma}$ are realized as 
hyperbolic M\"obius transformations of the Poincar\'{e} disk. What 
we have to know about hyperbolic M\"{o}bius transformations is that 
they each have two fixed points on the boundary of the disk, and 
that there is one and only one flow line of a hyperbolic M\"{o}bius 
transformation that is also a geodesic connecting the fixed points. 
(Elliptic M\"{o}bius transformations are not allowed in this context 
precisely because they always have a fixed point within the disk, 
and this would 
give rise to a conical singularity in the quotient space.) All that 
we have to do to ensure that ${\Gamma}$ consists of hyperbolic M\"obius 
transformations only is to choose a fundamental region such that it 
tessellates the disk and becomes a smooth manifold when its sides are 
identified---the identification of a pair of sides then defines the 
action of the generators of the group. 
Figure~\ref{fig4} should make it clear 
how we choose the fundamental region. The boundaries of this region must 
meet the boundary of the disk without crossing each other, and 
one may show that this condition is met whenever 
\begin{equation} \sinh^2{\frac{a}{2}} > 1 \ . \label{sinh} \end{equation}
 If this condition is not met \scri \ disappears and there 
is a conical singularity in the quotient space; in group theoretical 
terms the group ${\Gamma}$ would contain elliptic elements. \\
% Fig 4:
\parbox{\textwidth}{\begin{fig}{fig4}
  \mbox{\setlength{\epsfxsize}{.76\textwidth} \epsfbox{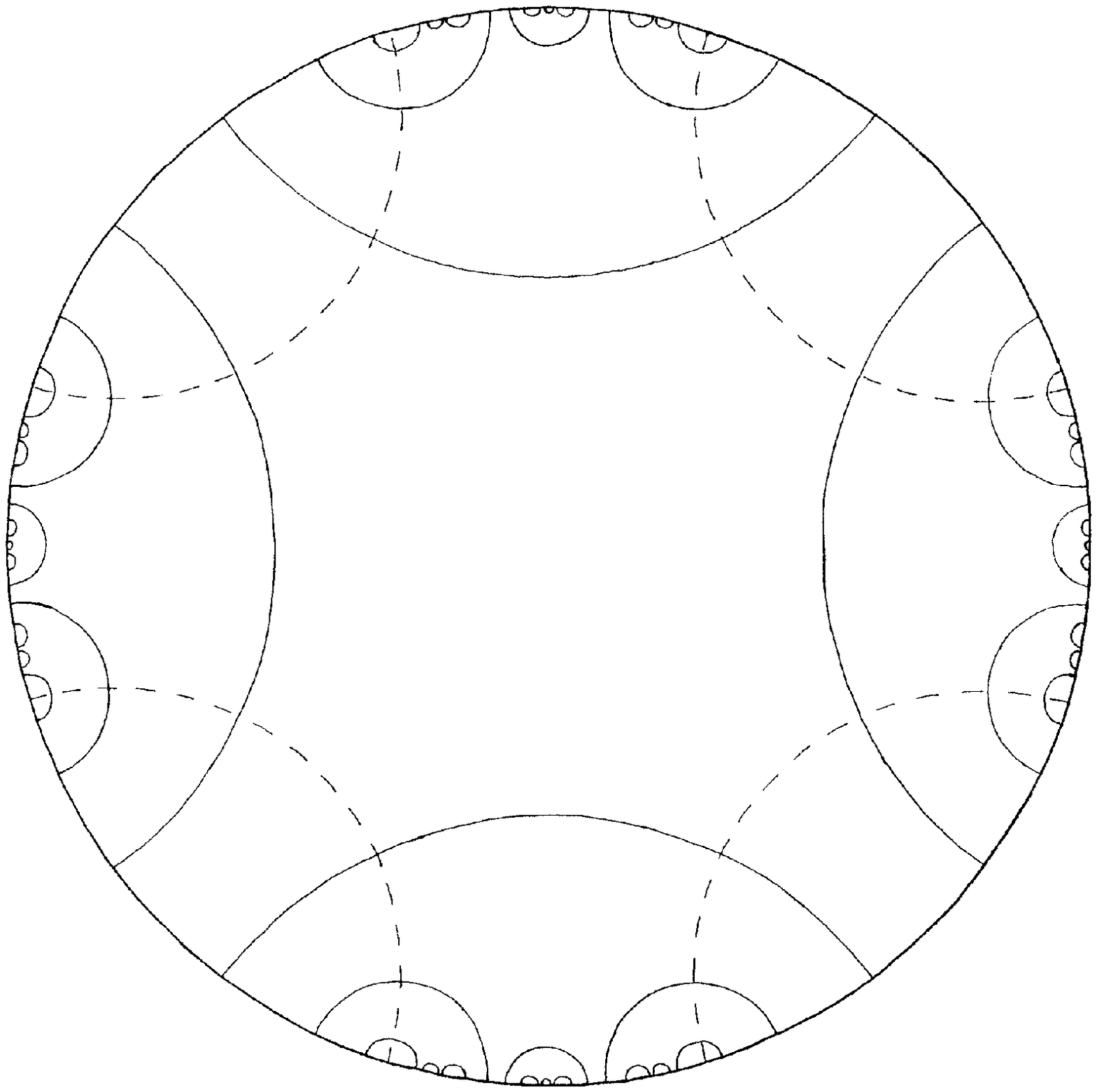} }\\[2 ex]
  \figcap{.90\textwidth}{  An "initial data surface" for a time
symmetric worm hole. It shows a fundamental region whose sides are
identified by ${\gamma}_1$ and ${\gamma}_2$, and how the disk is
tessellated by copies of this region. The geodesic flow lines of some
of the horizon words---namely (\ref{horisontord}) with
permutations---are shown as dashed lines. In the quotient space they
form the event horizon of the black hole.}
\end{fig}
}\\[2 ex]
\indent The covering space of the moment of time symmetry ($V = 0$) is the 
entire disk and the entire boundary minus the fixed points. Every 
element of ${\Gamma}$ contributes two fixed points, and every 
"word" in the generators is a group element. An important example is 
\begin{equation} 
\label{horisontord}
{\gamma}_{hor} = {\gamma}_1
{\gamma}_2{\gamma}^{-1}_1{\gamma}^{-1}_2 \ . 
\end{equation}
We call this word a horizon word,
since---as we will explain below---the unique geodesic connecting 
its fixed points lies in the event horizon of the black hole. The group
${\Gamma}$ contains an infinite set of copies of this word obtained 
by means of conjugation by arbitrary group elements ${\gamma}$:  
\begin{equation} 
{\gamma}^{-1}{\gamma}_{hor}{\gamma} = 
{\gamma}^{-1}{\gamma}_1
{\gamma}_2{\gamma}^{-1}_1{\gamma}^{-1}_2{\gamma} \ . 
\end{equation}
All these words, which include the permutations of the
group element (\ref{horisontord}), will also be called horizon words.
The Killing fields that generate them are all representatives of
one and the same field in the quotient space, namely, the generator of
rotations for the wormhole.

We need a good qualitative understanding of the set of all fixed 
points, or rather, the set of fixed point {\it free} segments of
$\scri$ at $t=0$. In order to obtain this set, begin by drawing the
fundamental region in the disk. The four segments of the 
boundary of the disk that belong to the fundamental region obviously
are free from fixed points. Next draw
all copies of the fundamental region that can be obtained by
transforming it with ${\gamma}_1$, ${\gamma}_2$ and their 
inverses. In this way a number of new fixed point free segments appear
as copies of the first ones. Namely, the four
fixed point free segments that we started out with are prolonged, and
eight additional such segments appear. Continue in this 
way, i.e. add further copies of the fundamental region until the 
entire disk has been tessellated, and in each step enlarge the set
with all the copies of the fixed point free segments obtained in the
preceding step. Note that this construction is reminiscent of the way
in which the Cantor dust is obtained, only that in this case we are
interested in the complement to the dust, rather than in the dust
itself. Two key observations are, first, that the original fixed point
free segments will grow so that they become segments ending in the
fixed points of the horizon words, and, second, that every other fixed
point free segment is situated between the fixed points of some
horizon word. The upshot of  
this discussion is that the covering space of the boundary at 
$t = 0$ can be described as the union of the fixed point 
free segments associated to the infinite set of horizon 
words. 

Having understood the "initial data" surface it is easy to understand 
the covering space of the \scri \ of the spinless wormhole. By definition 
this covering space is the intersection of all the regions where the 
flow of some element of ${\Gamma}$ is spacelike (all the flows are 
spacelike in the intersection), but a more useful 
description is to say that it is the union of a set of squares in 
one to one correspondence with all horizon word copies. Let 
us show this. We have 
a qualitative understanding of the location of the fixed points on 
the line $t = 0$. Each element in the group ${\Gamma}$ is associated 
with a unique pair of such fixed points. Moreover each such fixed 
point is the vertex of two lightlike lines along which the Killing 
vector generating this particular element of ${\Gamma}$ becomes 
lightlike. When we draw all these lines (in practice: some of 
them) we find that the boundary of the original 
anti-de Sitter space becomes divided into squares. Each fixed 
point free segment of the line $t = 0$ becomes the diagonal of 
a square in which the flow of every element of ${\Gamma}$ is 
spacelike. Hence the covering space of the \scri \ of the wormhole 
is given by the union of all such squares. But by the previous paragraph 
the fixed point free segments are in one to one correspondence 
with the copies of the horizon word, so this is what we wanted to show. 
There are four large squares 
lying between the fixed points of the original horizon words (that
is, (\ref{horisontord}) with permutations) and 
the future and past corners of these squares define the last and 
first points on \scri , respectively, see figure~5. It will be useful to 
quote the formula for the location of the last point on \scri . 
It occurs \cite{Fem} at $(t, {\phi}) = (t_P, {\phi}_P)$, where 
\begin{equation} \tan{t_P} = \sqrt{2\tanh^2{\frac{a}{2}} - 1} 
\hspace{15mm} {\phi}_P = \frac{\pi}{4} + \frac{n{\pi}}{2} \ . 
\end{equation}
%%
% Fig 5:
\parbox{\textwidth}{\begin{fig}{fig5}
  \mbox{\setlength{\epsfxsize}{.95\textwidth} \epsfbox{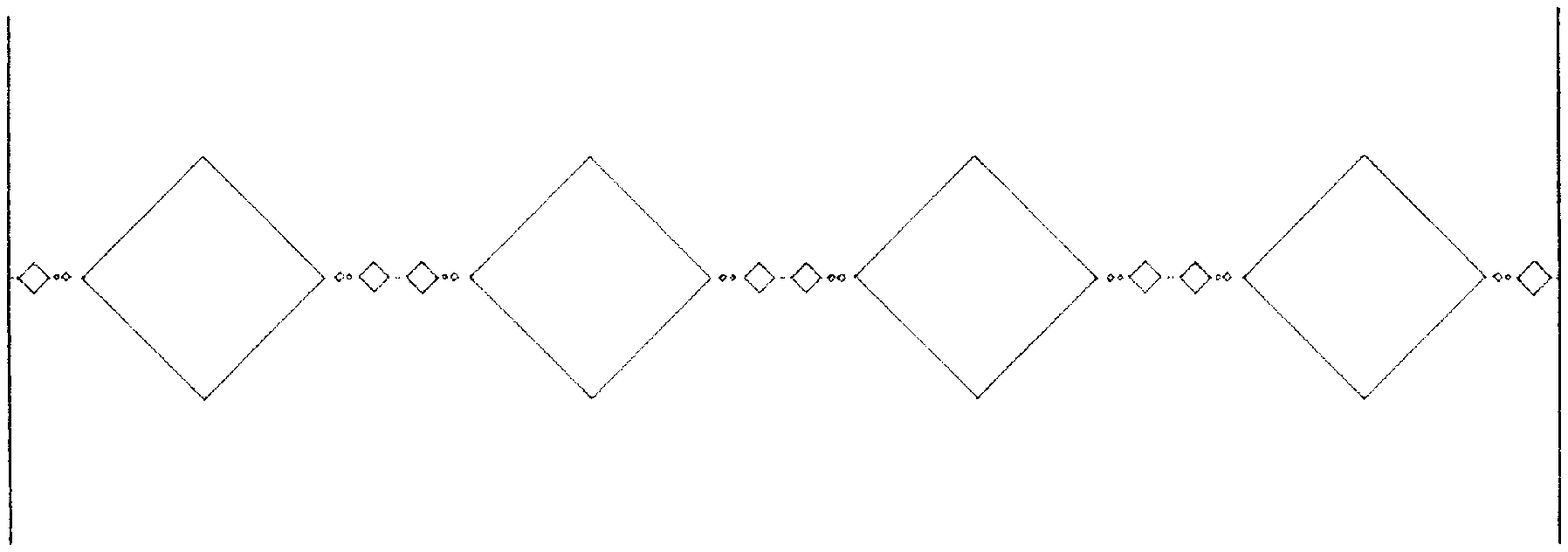} }\\[2 ex]
  \figcap{0.9\textwidth}{  This infinite set of squares on \scri \
(of which only a few are large enough to be visible in this figure)
is in one to one correspondence with all copies of the horizon word.
Namely, the corners on $t = 0$ of one of the squares are the two fixed
points of precisely one horizon word. Since there are no fixed points
of $\Gamma$ within the squares, all our Killing flows are spacelike
there. Thus the union of their interiors is the covering space of \scri
\ for the spinless wormhole.}
\end{fig}
}\\[2 ex]
More precisely there are four such points situated at 
90 degrees distance around the boundary; the first point on \scri \ 
has the sign of its $t$-coordinate reversed. Note that, 
appearances notwithstanding, the squares are isometric copies 
of each other---indeed we choose the scale of the metric at 
\scri \ so that this statement is true.

We will defer a discussion of the interior of the wormhole spacetime 
till section 7. However, let us give a few brief remarks that 
suffice to understand its domain of exterior communication. The 
basic point is that each of the squares on \scri \ is, in itself, 
simply an instance of the covering space of one asymptotic region 
of a BTZ black hole "in arbitrary position", as discussed in the 
previous section. By definition the 
covering space of the event horizon is the backwards light cone of 
the last point on \scri ; since this is a totally geodesic surface 
its intersection with the totally geodesic surface $t = 0$ is 
itself a geodesic, and it becomes a closed geodesic when the 
identification is carried through. Now we already know a closed 
geodesic in that homotopy class, namely the unique geodesic flow line 
of the generator of the horizon word, and therefore that flow line 
does indeed lie in the event horizon. There is another insight that 
comes for free at this point: Every square defines a bifurcate 
Killing horizon in the interior, defined by the backwards light cone 
from its future corner and the forwards lightcone from its past 
corner. The region in between \scri \ and this bifurcate Killing 
horizon is indeed isometric to the covering space of the domain 
of exterior communication of the BTZ black hole. This will remain 
true of the quotient spaces; hence the conclusion is that the 
domain of exterior communication of the time symmetric wormhole 
is isometric to that of the spinless BTZ black hole. Note that 
this does not \cite{Fem} mean that the non-trivial topology in 
the interior is unobservable from \scri . \\[1 cm]
{\bf 6. The Exterior of a Spinning Wormhole.}\\[5 mm]
Now we turn to the main point of this paper, the construction 
of a spinning wormhole. We concentrate on a simple and symmetric 
example where the Killing vectors are related by a rotation 
through 90 degrees. Thus we choose the type $I_b$ group elements 
\begin{equation} {\gamma}_1 = e^{{\xi}_1} \ , \hspace{10mm} 
{\xi}_1 = aJ_{XU} + bJ_{YV} \end{equation}
\begin{equation} {\gamma}_2 = e^{{\xi}_2} \ , \hspace{10mm} 
{\xi}_2 = aJ_{YU} - bJ_{XV} \ . \end{equation}
 We assume that $a > b \geq 0$. The group ${\Gamma}$ is 
defined as the free group formed from these generators. At first 
sight it may seem that we must now understand ${\Gamma}$ 
more or less "in the raw", since it will no longer be true that 
there is some particular Poincar\'{e} disk in anti-de Sitter space 
that is transformed into itself by all the elements of ${\Gamma}$. 
Nevertheless we will see that the general case is only marginally 
more difficult than the time symmetric case that we studied in 
the previous section. Let us first 
write the generating Killing vectors so that it becomes manifest 
how they act on \scri \ : 
\begin{equation} {\xi}_1 = - (a+b)J_1 - (a-b)\tilde{J}_1 = 
- (a+b)\sin{u}\partial_u - (a-b)\sin{v}\partial_v \end{equation}
\begin{equation} {\xi}_2 = (a+b)J_2 - (a-b)\tilde{J}_2 = 
- (a+b)\cos{u}\partial_u + (a-b)\cos{v}\partial_v \ . \end{equation} 
 The point to notice is that when we exponentiate the 
action of these generators then the transformation of the $u$ coordinate 
will be determined by the parameter $(a+b)$, while that of the 
$v$ coordinate will be determined by $(a-b)$. We may as well do this 
at once. The generators of ${\Gamma}$ act as projective transformations 
on the circles that are coordinatized by $u$ and $v$. It is convenient 
to introduce "rectifying" coordinates
\begin{equation} z = \tan{\frac{u}{2}} \hspace{13mm} \tilde{z} = 
\tan{\frac{v}{2}} \ . \end{equation}
 Then we find after a short calculation that 
\begin{equation} {\gamma}_1 = e^{{\xi}_1} \ : \hspace{21mm} 
z \rightarrow z' = \frac{e^{- \frac{a+b}{2}}z}{e^{\frac{a+b}{2}}} 
\hspace{20mm} \end{equation}
\begin{equation} {\gamma}_2 = e^{{\xi}_2} \ : \hspace{15mm} 
z \rightarrow z' = \frac{\cosh{\frac{a+b}{2}}z - \sinh{\frac{a+b}{2}}}
{- \sinh{\frac{a+b}{2}}z + \cosh{\frac{a+b}{2}}} \ , \end{equation}
 and similarly for the coordinate $\tilde{z}$ but with the 
parameter $(a+b)$ replaced with $(a-b)$. This is the action of the 
generators of the group ${\Gamma}$. To ensure that \scri \ 
exists we have to choose the parameters so that all the elements 
of ${\Gamma}$ belong to the conjugacy class $I_b = \mbox{hyp}\otimes \mbox{hyp}$. 
But we now see that this can be ensured by choosing 
the parameter $(a+b)$ so that all the projective transformations of 
the circle coordinatized by $u$ or $z$ are hyperbolic, and at the 
same time choosing 
the parameter $(a-b)$ so that the same is true for the circle 
coordinatized by $v$ or $\tilde{z}$. All that has happened is that 
we have two copies of the problem that we solved in the time symmetric 
case, and it follows that the conditions that ensure that \scri \ 
exists are 
\begin{equation} \sinh^2{\frac{a+b}{2}} > 1 \hspace{15mm} 
\sinh^2{\frac{a-b}{2}} > 1 \ . \end{equation}
 It remains to elucidate the properties of the 
resulting quotient space.

A qualitative understanding of \scri \ is in fact already 
in hand; on the line $t = 0$ we mark the fixed points of a 
spinless wormhole with parameter $(a+b)$ and draw lines of constant 
$u$ through these points. Then we change the parameter to $(a-b)$, 
mark the fixed points, and draw lines of constant $v$ through 
those. The fixed points of the spinning wormhole will occur where 
two such lines belonging to the same group element meet. In the 
spinless case the set of fixed points lying on the line $t = 0$ 
played a special role. As we increase the parameter $b$ these 
fixed points move away from this particular line, but it is possible 
to show that they still lie on a spacelike (but quite "jagged") line. 

Indeed the whole picture is very similar to what it was in the 
spinless case, the only difference is that what used to be squares 
where the flow of all the Killing vectors is spacelike has become 
rectangles,as shown in figure~\ref{fig6}. The rectangles are isometric copies of each other 
and each rectangle can be regarded as an instance of 
the covering space of \scri \ for a BTZ black hole "in general 
position". Moreover each rectangle defines a bifurcate Killing 
horizon in the interior and the domain of exterior communication 
of our wormhole is
covered by the union of all such domains, one for each rectangle. 
Hence it is isometric to the domain of exterior communication 
for a single BTZ black hole. The event horizon is given by the 
backwards lightcone of the last point on \scri , as usual. 
When we go to the quotient space only one asymptotic region will 
remain (as can be shown by choosing a fundamental region on 
\scri ).\\
% Fig 6:
\parbox{\textwidth}{\begin{fig}{fig6}
  \mbox{\setlength{\epsfxsize}{.95\textwidth} \epsfbox{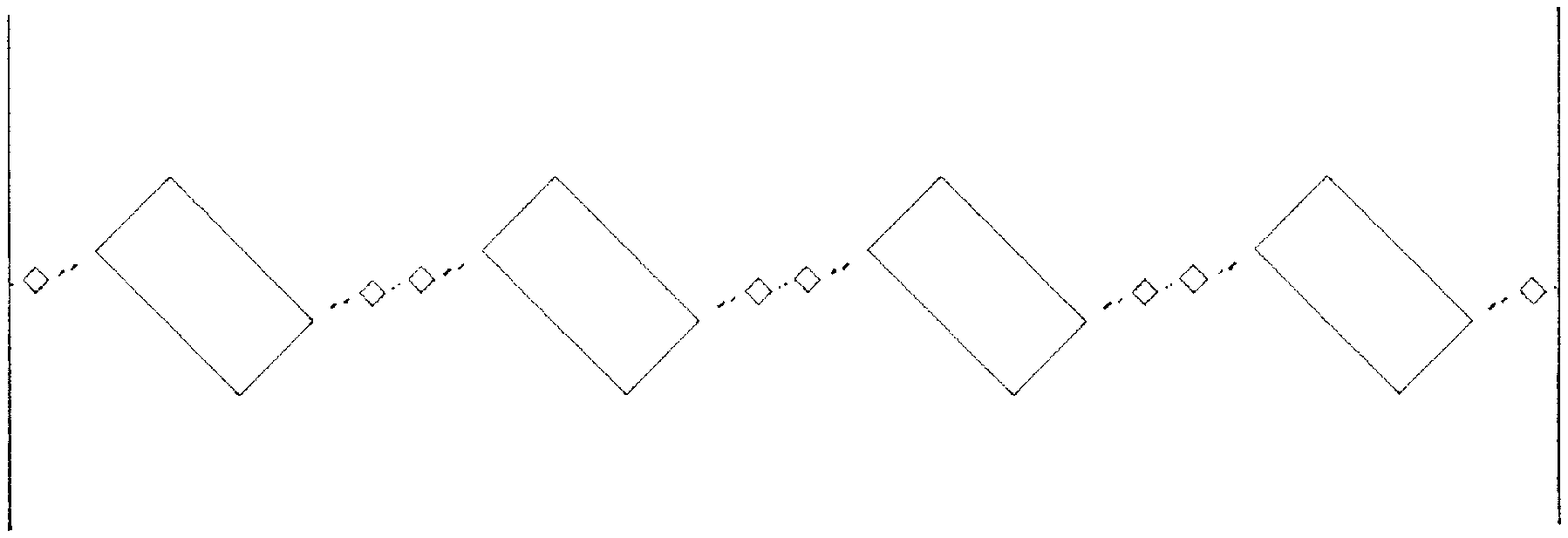} }\\[2 ex]
  \figcap{0.9\textwidth}{  This is the analogue of figure 5 for the
spinning case: the union of the rectangles shown is the covering
space of \scri \ for the spinning wormhole. The corners of the large
rectangles are the fixed points of the horizon word
(\ref{horisontord}) with permutations, and the first and last 
points on \scri . The fixed points that in the spinless case occurred
on the line $t = 0$ now lie on spacelike (but quite "jagged") lines
connecting the fixed points of all the horizon words.}
\end{fig}
}\\[2 ex]
\indent We want to compute the angular velocity of the horizon 
with respect to \scri . This is a considerably messy calculation. 
In logical outline what we have to do is to compute the generator 
of rotation from the equation
\begin{equation} e^{{\xi}_{rot}} = 
{\gamma}_1{\gamma}_2{\gamma}_1^{-1}{\gamma}_2^{-1} \ . \end{equation} 
 Then we compare the result to our previous general form 
of a type $I_b$ Killing vector, eq. (\ref{xi}), and read off the 
parameter $k$. This parameter together with the known values for 
the light cone coordinates of the fixed points that terminate \scri \ 
to the future and past can then be inserted into our general formula 
for the angular velocity, eq. (\ref{Omega}). We were unable to 
simplify this calculation very much. Anyway, the result is 
\begin{equation} {\Omega} = \frac{c - \tilde{c}}{c + \tilde{c}} \ , 
\end{equation}
 where 
\begin{equation} c = - 2\sinh^2{\left( \frac{a+b}{2}\right)}\sqrt
{\sinh^4{\left( \frac{a+b}{2}\right)} - 1} \ \ \end{equation}
 and 
\begin{equation} \tilde{c} = - 2\sinh^2{\left( \frac{a-b}{2}\right)}\sqrt
{\sinh^4{\left( \frac{a-b}{2}\right)} - 1} \ . \end{equation}
 These formul\ae \ provide the desired quantitative 
expression for the rate of spin of our wormhole.

It remains to count the number of parameters in our solution. To do so we 
observe that six parameters are needed to characterize an element 
of the six dimensional isometry group $SO(2,2)$. The discrete 
group ${\Gamma}$ is generated by two such elements, and we must 
remember that the final spacetime is invariant under a global 
isometry described by six parameters. Hence the number of 
parameters in our solution is $2\times 6 - 6 = 6$. In the 
time symmetric case the group elements are to be chosen in 
the three dimensional group $SO(2,1)$, so that the number of 
parameters in the time symmetric case is only $2\times 3 -3 = 3$.
 In effect we are using the analogue of Fricke-Klein 
coordinates on the parameter space, while in our previous 
publication \cite{Fem} we used Fenchel-Nielsen coordinates. 
As we will see in the next section the spatial topology of 
our wormhole is that of a torus with an asymptotic region 
attached. The generalization to the higher genus case is 
straightforward in principle. 

As an aside, we note that it is rather difficult to understand 
the parameter spaces in terms of the kind of presentation of 
the group that we have been employing here. Consider the 
time symmetric case for simplicity. The three parameters are 
the parameters $a$ and $a'$ multiplying the Killing 
vectors that are to be exponentiated to get the generators, 
and the angle ${\alpha}$ between their geodesic flow lines 
in the initial data disk. The condition on these parameters 
that guarantee that a black hole solution results is 
\begin{equation} \coth^2{\frac{a}{2}} + \coth^2{\frac{a'}{2}} 
- \coth^2{\frac{a}{2}}\coth^2{\frac{a'}{2}} > \cos^2{\alpha} \ . 
\end{equation}
 When $a = a'$ and ${\alpha} = {\pi}/2$ this reduces 
to eq. (\ref{sinh}).\\[1 cm]
{\bf 7. The Interior of the Spinning Wormhole.}\\[5 mm]
It is time to discuss the interior of our spinning 
wormhole. It is {\it a priori} not so clear 
whether---like the spinning BTZ black hole as well as the 
ordinary Kerr solution---our spinning wormhole has a mouth 
behind the event horizon through which an observer may travel 
into a new universe, identical with the one he left. As a 
matter of fact this is not the case. To see this we have to 
understand the covering space of the interior.

First recall the situation for the non-rotating wormhole. In that 
case all the elements of ${\Gamma}$ act as hyperbolic M\"{o}bius 
transformations in the Poincar\'{e} disk at $t = 0$. In particular 
they all have two fixed points at its boundary. As may be shown, the 
"singularity surface" belonging to such an element consists of 
lightlike surfaces growing up forwards and backwards in time 
from each of its two fixed points at $t = 0$. In other words, 
they are light cones with their apices at \scri . The covering 
space---by definition a region where all the flows are 
spacelike---is then easily visualized. Just locate the fixed 
points at $t = 0$ of all the elements in ${\Gamma}$, and remove 
all points in the causal past and the causal future of 
these fixed points. What remains is the covering space, which we have
tried to depict in figure~\ref{fig7}. It lasts 
only for a finite amount of coordinate time, 

that is until all these light cones meet in the interior. In our coordinates 
this will happen at $t = \pm {\pi}/2$ and ${\rho} = 0$.

In the spinning case things get more involved because the 
"singularity surfaces" are timelike. In the case of 
the spinning BTZ black hole this has the consequence that 
a mouth opens up in the disk at $t = {\pi}/2$ through 
which an observer may travel to another universe, identical 
with the one she left. We will show that this does not happen 
in the spinning wormhole. Instead it turns out that there 
are certain elements of the group ${\Gamma}$ whose "singularity 
surfaces" get arbitrarily close to null planes, i.e. lightcones 
with their apices at infinity. Moreover there is such an apex 
arbitrarily close to any given fixed point on the jagged 
spacelike line where---according to section 6---the fixed 
points lie. Thus the covering space can be visualized in 
much the same way as in the non-rotating case, that is as what 
is left when all points to the future and to the past of 
these fixed points have been removed. 

Since we will consider families of timelike surfaces that have a 
null surface as a limit we should make clear that this notion 
makes sense because we regard the coordinate system as fixed 
in the discussion. The idea is that all points on one side of 
any surface in the family are to be removed from covering space (since 
the Killing flow is timelike there). If it is the case that 
any given point on one side of a null surface is such that a 
member of the family of timelike surfaces passes between 
the point and the null surface, then the family is said to 
have the null surface as a limit, and the given point must 
be removed from covering space. In effect the null surface 
can be regarded as a "singularity surface", except that points 
on the null surface itself are not removed from covering space.\\
% Fig 7:
\parbox{\textwidth}{\begin{fig}{fig7}
  \mbox{\setlength{\epsfysize}{.40\textheight} \epsfbox{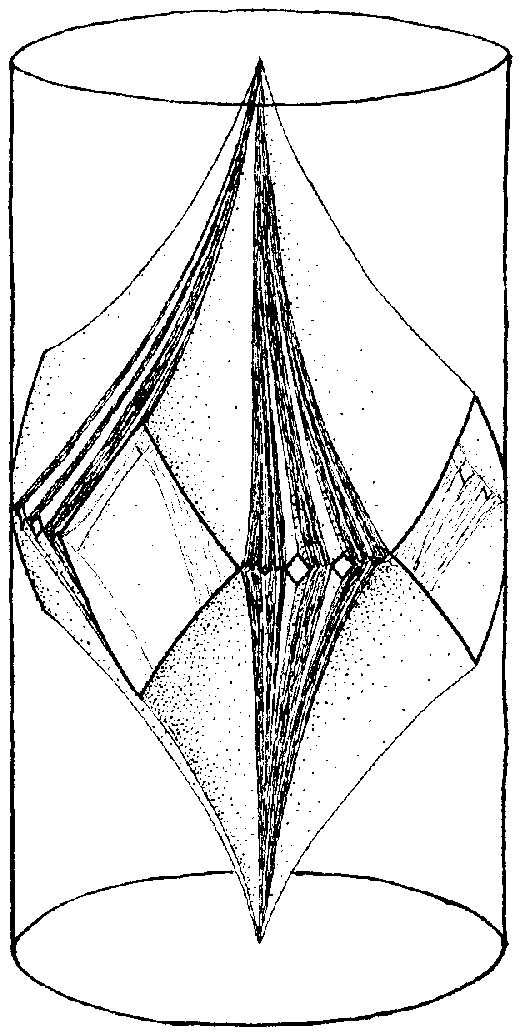} }\\[2 ex]
  \figcap{.9\textwidth}{  Here we have tried to visualize the full
covering space of the spinless wormhole in the anti-de Sitter
cylinder. Thus figure 5, which shows its covering space on \scri \
only, is simply the surface of this cylinder. Indeed, two of the four
large squares of that figure are visible also here, while the other two
are hidden behind the structure. Formally, to obtain the boundary of
the full covering space from figure 5 we should evolve lightcones
forward and backward from the fixpoints at $t = 0$ of all the elements
in $\Gamma$. Since it takes a time $\pi / 2$ for a lightray to travel
from infinity to the centre of the cylinder, the resulting diamond
shaped structure will last between $t=-\pi/2$ to $t=+\pi/2$.
}
\end{fig}
}\\[2 ex]
\indent Let us first show how this works for the fixed points of the 
generator ${\gamma}_2$. Consider the group element 
\begin{equation} {\gamma}_m \equiv {\gamma}_2^m{\gamma}_1
{\gamma}_2^{-m} \ . \end{equation}
 This is simply the generator ${\gamma}_1$ transformed 
$m$ times by ${\gamma}_2$. Somewhat lengthy calculations shows 
that this group element is obtained by exponentiating the 
Killing vector
\begin{eqnarray} {\xi}_m = - (a+b)( J_1\cosh{m(a+b)} + 
J_3\sinh{m(a+b)} ) - \nonumber \\
\ \\
\hspace{22mm} - (a-b)(\tilde{J}_1\cosh{m(a-b)} - \tilde{J}_3
\sinh{m(a-b)}) \ . \nonumber \end{eqnarray}
 The norm of ${\xi}_m$ becomes
\begin{eqnarray}
 \|{\xi}_m\|^2 = \frac{a^2 + b^2}{2} + 
\frac{a^2 - b^2}{4}\left( e^{2ma}(Y-U)^2 \right. + \nonumber \\
\ \\
\left. + e^{-2ma}(Y+U)^2 - e^{2mb}(X+V)^2 - e^{-2mb}(X-V)^2 \right)
\ . \nonumber 
\end{eqnarray}
 Now let us take the limit $m \rightarrow \infty $. 
If $Y \neq U$ we get (since $a > b$) 

\begin{equation} \lim_{m \rightarrow \infty } \|{\xi}_m\|^2 = 
\frac{a^2 - b^2}{4}e^{2ma}(Y - U)^2 \ . \label{Soren} \end{equation}

 The "singularity surface" of the element ${\gamma}_m$ 
is defined as the surface where ${\xi}_m$ is null, and it 
coincides with the "singularity surface" of ${\gamma}_1$ transformed 
$m$ times by ${\gamma}_2$. What we have shown is that in the limit 
$m \rightarrow \infty $ this surface approaches the surface 
$Y = U$, which is a light cone with its vertex at one of the 
fixed points of ${\gamma}_2$. Note however that eq. (\ref{Soren}) 
fails on this surface; if we set $Y = U$ before we take the limit 
we see that the ${\xi}_m$ is actually a timelike vector on this 
surface. 

We can already conclude that there can be no mouth inside the 
spinning wormhole, because the "singularity surfaces" of ${\gamma}_m$ 
and its inverse for large enough $m$ get arbitrarily close to 
lightcones with apices at the fixed points of ${\gamma}_2$. These 
meet each other at $t = {\pi}/2$, so if the covering space does 
not end before that it certainly does so then. 

We would also like to show that there are no inner horizons. To 
do this we have to show that the boundary of covering space is 
everywhere lightlike rather than timelike, that is to say that 
we have to show that there is a "singularity surface" 
that is arbitrarily close to a light cone having its apex 
at an arbitrary given fixed point. We will offer a heuristic argument 
to this effect. Consider the "singularity surface" of the word 
\begin{equation} {\eta}^m{\gamma}{\eta}^{-m} \ , \end{equation}
 where ${\eta}$ and ${\gamma}$ are arbitrary elements 
of ${\Gamma}$. The claim is that for large $m$ this surface 
approaches a light cone with its apex at one of the fixed 
points of ${\eta}$. 

Let ${\eta} = e^{\xi}$ and let $S_m$ be the timelike "singularity 
surface" of ${\gamma}$ transformed $m$ times with ${\eta}$. We 
assume that ${\xi}$ is never tangent to $S_0$, so that indeed 
the entire surface is transformed by ${\eta}$. That is, we 
assume that 
\begin{equation} \frac{\xi}{\|{\xi}\|}\cdot n_0 \neq 0 \ , 
\end{equation}
 where $n_0$ is the normal of $S_0$. But the normal 
$n_m$ to the surface $S_m$ is defined as the transformation 
of $n_0$ along the Killing flow of ${\xi}$ , and therefore 
it must be true that 
\begin{equation} \frac{\xi}{\|{\xi}\|}\cdot n_m = 
\frac{\xi}{\|{\xi}\|}\cdot n_0 . \end{equation} 
 Now we want to know what happens to $S_m$ as $m$ 
goes to infinity. We assume that it becomes a limiting surface 
$S_{\infty}$. But since ${\xi}$ has a component along the normal 
of the surfaces for all $m$ the limiting surface can exist only 
if the limiting normal lies in the limiting surface, that is 
to say that $S_{\infty}$ must be a null surface. Since its apex 
on \scri \ is being pushed by ${\eta}$ towards one of its own 
fixed points it follows that for large enough $m$ the "singularity 
surface" $S_m$ will come arbitrarily close to a light cone with 
its apex at one of the fixed points of the arbitrary group element 
${\eta}$. This proves our claim, provided that the assumptions 
that we made are accepted. Hence the covering space of the 
spinning wormhole can be visualized in the same manner as the 
covering space of the spinless wormhole, and there can be 
no inner horizons.

Finally, the topology of the quotient space: Is it a wormhole? 
Yes. According to the previous section there is only one asymptotic 
region attached. Moreover the first homotopy group of the quotient 
space is equal to ${\Gamma}$, just as it was in the case of 
the spinless wormhole (even though the details of how ${\Gamma}$ 
acts on the covering space have been changed). Hence the topology 
of the quotient space is the same in both cases, namely that 
of a torus with an asymptotic region attached. Note that higher genus 
wormholes can be constructed if they are wanted.\\[1 cm]
\noindent {\bf 8. Conclusions and Open Questions.} \\[5 mm]
Our conclusions are simple to state. We have constructed 
a spacetime of the form $adS/{\Gamma}$ whose spatial topology 
is that of a torus with an asymptotic region attached. The 
domain of exterior communication is isometric to that of 
the spinning BTZ black hole \cite{BHTZ}, which means that there is 
an event horizon that becomes stationary at late times and 
that spins with respect to infinity. All "singularities" in 
the interior are hidden behind this event horizon. The 
interior ends in a "singularity" and there are no inner 
horizons. The solution is a member of a six parameter 
family of solutions having similar properties (and which 
includes a three parameter subfamily of spinless wormholes). 

Our argument for the absence of inner horizon was heuristic 
and ought to be tightened up. This should be doable. 
To see what the open questions are, let us remind ourselves about 
the spinless wormholes \cite{Fem}. The parameter space of 
all such solutions is well understood for arbitrary genus. Their 
domain of exterior communication is isometric to that of the 
spinless BTZ black hole. We also have a detailed description of the 
topology of the event horizon (including its caustics), the 
location of all the marginal trapped surfaces is known, and it is 
understood in what sense active---but not passive---topological 
censorship holds. Clearly it would be of interest to 
bring our understanding of the spinning case up to the same level, 
so for the spinning case this becomes a list of open question; 
we expect that they should prove easy to settle. A different 
question that should be addressed at the classical level is 
the dynamics of these spacetimes from a Hamiltonian point of 
view. The case of compact locally anti-de Sitter spacetimes 
is well understood \cite{Lars}; the non-compact case considered 
here offers complications of some interest.

In spite of the loose ends we think that it is fair to say that 
the Clifford-Klein problem that we have posed is solved. There 
is a similar problem in 3+1 dimensions: Find all suitably 
regular 3+1 dimensional spacetimes of the form $adS/{\Gamma}$. 
Natural generalizations of the BTZ black holes do exist \cite{Fyra}. 
These solutions have been explored and extended in various 
directions \cite{Tre} \cite{Mann}, but it is conceivable that 
further solutions of interest may exist. 

Beyond these classical questions there looms the question whether 
our wormholes can provide soluble examples that may help us to 
understand the quantum mechanics of black holes and topological 
geons in general. Whether this is so remains to be seen.\\[1 cm]
{\bf Acknowledgement.} \\[5 mm]
Ingemar Bengtsson was supported by NFR.

\end{document}